Scattering

\documentclass[12pt]{article} \input epsf.tex

\newcommand{\be}{\begin{equation}}
\newcommand{\ee}{\end{equation}}
\newcommand{\bear}{\begin{eqnarray}}
\newcommand{\eear}{\end{eqnarray}} \newcommand{\ba}{\begin{array}}
\newcommand{\ea}{\end{array}}

\newcommand{\CF}{{\cal F}}
\begin{document}

\pagestyle{empty} \begin{titlepage}
\def\thepage {} 

\title{\Large \bf 
Analytic Estimates of the QCD Corrections to Neutrino-Nucleus
Scattering
}

\author{\normalsize
\bf \hspace*{-.3cm} 
Bogdan A.~Dobrescu and R.~Keith Ellis
\\ \\ 
{\small {\it
Theoretical Physics Department, Fermilab, Batavia, IL 60510, USA }}\\
}

\date{ } \maketitle

\vspace*{-7.4cm}
\noindent \makebox[11cm][l]{\small \hspace*{-.2cm} October 13, 2003} 
{\small FERMILAB-Pub-03/304-T} \\
\makebox[11cm][l]{\small \hspace*{-.2cm} hep-ph/0310154} 
{} \\
{\small } \\

\vspace*{7.9cm}

\begin{abstract}
{\small
We study the QCD corrections to neutrino
deep-inelastic scattering on a nucleus, and
analytically estimate their size.
For an isoscalar target, we show that the 
dominant QCD corrections to the 
ratio of the neutral- to charged-current 
events  are suppressed by $\sin^4\theta_W$, where 
$\theta_W$ is the weak mixing angle.
We then discuss the implications for the 
NuTeV determination of $\sin^2\theta_W$.
}

\end{abstract}

\vfill \end{titlepage}

\baselineskip=18pt \pagestyle{plain} \setcounter{page}{1}

\section{Introduction}

For more than three decades, 
neutrino deep-inelastic scattering has been an essential 
source of information regarding both the electroweak interactions
and the structure of the nucleons. 
A very important quantity measured in neutrino (antineutrino)
deep-inelastic scattering 
is the ratio $R^\nu$ ($R^{\overline{\nu}}$)
of the total cross sections for the 
neutral- and charged-current processes.
The most precise measurements to date of $R^\nu$
and $R^{\overline{\nu}}$
have been performed by the NuTeV collaboration
\cite{Zeller:2001hh}, which led to a determination of 
$\sin^2\theta_W$ ($\theta_W$ is the weak mixing angle)
with uncertainty of less than a percent.
Such a precision makes
the inclusion of QCD corrections a necessary part of 
the determination of $\sin^2\theta_W$.

The next-to-leading order (NLO) QCD corrections, 
{\it i.e.}, of order $\alpha_s$, 
to neutrino-nucleon cross sections 
have been known for a long time \cite{Bardeen:1978yd,
Altarelli:1978id, Furmanski:1981cw}, 
and the order $\alpha_s^2$ corrections have also been computed
\cite{Zijlstra:1992qd, Zijlstra:1992kj, Kazakov:fu}. However, to our 
knowledge, a careful analysis of the size 
of even the NLO QCD corrections to $R^\nu$
and $R^{\overline{\nu}}$ has not yet been performed.
Part of the reason is the observation that 
the NLO QCD corrections to the Paschos-Wolfenstein 
ratio of differences of cross sections \cite{Paschos:1972kj}, 
$R_{\rm PW}$,
cancel for an isoscalar target \cite{LlewellynSmith:ie}. 
Most discussions of perturbative QCD corrections
to the NuTeV determination of $\sin^2\theta_W$ have been
concentrated on $R_{\rm PW}$ \cite{Davidson:2001ji, Kulagin:2003wz,
McFarland:2003jw}. 
However, the relation between the NLO QCD corrections to $R_{\rm PW}$
and those to $R^\nu$ and $R^{\overline{\nu}}$ is not clear.
In fact, it has been often claimed that 
the NLO QCD corrections to $R^\nu$ and $R^{\overline{\nu}}$
are expected to be as large as 10\% (see \cite{Davidson:2001ji,
Gambino:2002xp, Davidson:2002fb, Strumia:2003pc}),
given that the expansion parameter of the perturbative series 
is typically $\alpha_s/\pi$, where $\alpha_s$ is 
evaluated at a scale of about 20 GeV$^2$. 
The NuTeV analysis takes into account a variety of corrections 
to the cross sections, including a partial, phenomenological 
description of the QCD corrections. However, the latter 
might differ from the result of a systematic expansion in  
$\alpha_s$, and therefore it is essential to know 
how large these corrections are.

In this paper we derive an analytic, approximate expression
for the NLO QCD corrections to $R^\nu$ and $R^{\overline{\nu}}$. 
We show that these are suppressed by an additional factor of 
$\sin^4\theta_W$.  
This conclusion is consistent from an order-of-magnitude point of view
with the numerical results presented in
Ref.~\cite{Kretzer:2003iu}. We then
address the issue of how these corrections might change 
the NuTeV result for $\sin^2 \theta_W$. A definitive statement will 
require a re-analysis including full NLO effects 
by the NuTeV collaboration. 

We emphasize that there are several kinds of QCD corrections
that may affect the NuTeV analysis.
First there are perturbative QCD corrections to the 
differential cross section, which are computable 
in the standard model, and are the focus of this paper.
Second, there are nonperturbative effects, such as 
higher twist effects, which have been included in the NuTeV 
analysis (see section 5.1.12 of \cite{Zeller:2002he}).
Third, there are corrections to the parton distribution functions
(PDF's), which are being studied by various groups
\cite{Zeller:2002du, Martin:2003sk, cteq}, and are 
 not discussed here.

In Section 2 we review the lowest order differential cross section,
and in Section 3 we present the order $\alpha_s$ corrections to the 
differential cross section. 
We then integrate (in Section 4) the differential cross section and use 
(in Section 5) some perturbative expansions to obtain 
analytical expressions for the order $\alpha_s$ corrections
to $R^\nu$ and $R^{\overline{\nu}}$. We
estimate in Section 6 the impact of the perturbative QCD corrections  
on the determination of $\sin^2\theta_W$, and we comment on our 
results in Section 7.

\section{$\nu$-nucleus cross section at leading order}
\setcounter{equation}{0}

We consider neutrino deep-inelastic scattering on a nucleus,
ignoring the Fermi motion of the nucleons.
In the lab frame, the inclusive $\nu_\mu$-nucleus collision is described by 
three kinematic variables: the squared momentum transfer, $Q^2$, 
the energy $E_\nu$ of the incoming neutrino,
and the inelasticity parameter $y$, which is 
the fraction of the lepton energy lost in the lab frame.
In the  parton model, $Q^2$ may be expressed in terms of 
the fraction $x$ of the nucleon momentum, averaged over the entire nucleus:
\be 
Q^2  = 2xy M_N E_\nu  ~.
\label{qsquare}
\ee
Here $M_N$ is the average nucleon mass in the nucleus, we are neglecting 
the parton mass, and both $x$ and $y$ range from 0 to 1.

To be specific, we will concentrate on an iron nucleus, but our considerations 
apply to any target which is approximately isoscalar.
Neglecting the muon mass, there are three 
structure functions that contribute to 
the $\nu_\mu$-nucleon differential
cross sections in the lab frame:
\be
\frac{d\sigma^{\rm C,N}\left( \nu_\mu {\rm Fe} \right)}{dx dy}
= \frac{x M_N E_\nu G_{\rm F}^2}{\pi \left(1 + Q^2/M_{W,Z}^2\right)^2 }
\left[ \frac{y^2}{2} \CF_1^{\rm C,N}
+ \left( 1 - y - \frac{xy M_N}{2 E_\nu } \right) \CF_2^{\rm C,N}
+ y \left( 1 - \frac{y}{2} \right) \CF_3^{\rm C,N}
\right] ~,
\label{differential}
\ee
where $G_{\rm F}$ is the Fermi constant, and the 
inclusive cross sections for the charged- and neutral-current processes,
$\nu_\mu {\rm Fe} \rightarrow \mu^- X$ and 
$\nu_\mu {\rm Fe} \rightarrow \nu_\mu X$, are labeled respectively
by $\sigma^{\rm C}\!\left( \nu_\mu {\rm Fe}\right)$ 
and $\sigma^{\rm N}\!\left( \nu_\mu {\rm Fe}\right)$.
Note that instead of the structure functions introduced here, 
$\CF_i \equiv \CF_i(x,Q^2)$ with $i=1,2,3$, which are convenient 
for the discussion of NLO corrections, the textbooks typically use 
$F_1 = \CF_1/2$, $F_2 = x \CF_2$, $F_3 = \CF_3$.

The structure functions can be written as expansions in several small parameters,
\be
\CF_i^{\rm C,N} = \CF_{i\, {\rm LO}}^{\rm C,N} + \delta \CF_i^{\rm C,N}
+ O\left(\frac{\alpha}{\pi\sin^2\theta_W}\right) 
+ O\left(\frac{M_N^2}{Q^2}\right) + O\left(\frac{m_c^2}{Q^2}\right) + ... ~.
\label{corr}
\ee
The first term of the expansion is due to a $W$ or $Z$ exchange without any 
radiative corrections and in the limit where the momentum transfer is much larger 
than the mass of any particle in the initial or final state.
For the charged-current process,
\bear &&
\CF_{1\, {\rm LO}}^{\rm C} = \CF_{2\, {\rm LO}}^{\rm C} 
= 2 \left( d + s + \overline{u} + \overline{c} \right)
\nonumber \\ [1em] && 
\CF_{3\, {\rm LO}}^{\rm C} = 2 \left( d + s - \overline{u} - \overline{c} \right) ~.
\label{CC-structure-functions}
\eear
where $q \equiv q(x, Q^2)$, with $q=u,d, s, c$, 
is the probability distribution, averaged over the entire nucleus, 
for finding the parton $q$ with momentum fraction $x$ inside a
nucleon of the iron nucleus, when the squared momentum transfer is 
$Q^2$. 

We have included only quarks of the lighter two generations, 
because for the $b$ quark the PDF is sufficiently 
small to be neglected at the NuTeV energies, and the deviations 
from unitarity of the diagonal block of the CKM matrix associated with 
the first two generations are of order $10^{-3}$ ($|V_{ts}|^2$ or
$|V_{cb}|^2$).

The leading-order structure functions for the neutral-current process are
\bear &&
\CF_{1\, {\rm LO}}^{\rm N} = \CF_{2\, {\rm LO}}^{\rm N} = 
2 \left( {g_L^u}^2 + {g_R^u}^2 \right) 
\left( u + c + \overline{u} + \overline{c} \right)
+ 2 \left( {g_L^d}^2 + {g_R^d}^2 \right) 
\left( d + s + \overline{d} + \overline{s} \right)
\nonumber \\ [1em] && 
\CF_{3\, {\rm LO}}^{\rm N} = 2 \left( {g_L^u}^2 - {g_R^u}^2  \right) 
\left( u + c - \overline{u} -
\overline{c} \right)
+ 2\left( {g_L^d}^2 - {g_R^d}^2  \right) 
\left( d + s - \overline{d} - \overline{s} \right) ~.
\label{NC-structure-functions}
\eear
As usual, $g_L^{u,d}, g_R^{u,d}$ are the quark couplings 
to the weak bosons, which depend on the electric charge,
$Q^{u,d}$, and on the weak mixing angle, $\theta_W$:
\bear &&
g_L^{u,d} = \pm\frac{1}{2} - Q^{u,d} \sin^2\theta_W \; , 
\nonumber \\ [.4em] &&
g_R^{u,d} =  -  Q^{u,d}\sin^2\theta_W \; .
\eear
The $\overline{\nu}_\mu$-nucleus differential cross sections are obtained from the 
$\nu_\mu$-nucleus ones by interchanging the $q$ and $\overline{q}$ distributions.

The term $\delta \CF_i^{\rm C,N}$ in Eq.~(\ref{corr})
represents the NLO QCD corrections, and is of order
$O\left(\alpha_s/\pi\right)$, where 
$\alpha_s(Q^2) \approx 0.2$ for the average momentum transfer at
NuTeV. Therefore, these corrections are a priori expected to 
be large, and their impact on the ratios of neutral- to charged-current 
events, $R^\nu, R^{\bar{\nu}}$, are the focus of this paper.

The electroweak corrections, encoded in the
third term of the expansion (\ref{corr}), come from 
loops involving electroweak gauge bosons, the top quark,
and the Higgs boson, as well as 
from the emission of a real photon. The photon corrections, although 
not enhanced by a $1/\sin^2\theta_W$ factor,
turn out to be dominant because 
their contributions to the charged- and neutral-current processes 
are substantially different, and lead to a shift of a few percent
in the values of $R^\nu$ and $R^{\bar{\nu}}$ at NuTeV \cite{Zeller:2002he}.
The target mass corrections,
are of order $M_N^2/Q^2 \approx M_N/E_\nu$, so that we 
expect them to be at most as large as a few percent.
A recent discussion of the target mass corrections
is given in Ref.~\cite{Kretzer:2003iu}.
The charm mass affects mainly the charged-current scattering off the 
strange sea, and accounts for a shift of about 2\% in $R^\nu$ and 
$R^{\bar{\nu}}$ \cite{Zeller:2002he}.
Details of how all the above corrections have been included in the NuTeV 
analysis can be found in Ref.~\cite{Zeller:2002he}.

\section{Next-to-Leading Order QCD Corrections to the $\nu$-Fe 
Differential Cross Sections}

\setcounter{equation}{0}

It is convenient to compute 
the QCD corrections to the parton-level cross sections 
in the DIS scheme, where only the $\CF_1$ and $\CF_3$ structure functions
change \cite{Altarelli:1978id}.

The NLO QCD corrections to the $\CF_1$ structure functions 
are due to one-loop contributions involving a gluon, and  from 
the emission or absorption of a real gluon, which includes scattering off the 
gluon sea: 
\bear 
&&
\delta \CF_1^{\rm C} = 
- \frac{4\alpha_s}{3\pi} \int_x^1 dz
\left[ \CF_{1\, {\rm LO}}^{\rm C}\left(\frac{x}{z}, Q^2\right) + 
6 (1-z) g\left(\frac{x}{z}, Q^2\right) 
\right] ~,
\nonumber \\ [1em] &&
\delta \CF_1^{\rm N} = 
- \frac{4\alpha_s}{3\pi} \int_x^1 dz
\left[ \CF_{1\, {\rm LO}}^{\rm N}\left(\frac{x}{z}, Q^2\right) + 
6\left(g_L^2 + g_R^2\right) (1-z) g\left(\frac{x}{z}, Q^2\right) 
\right] ~,
\label{f1-nlo}
\eear
where $g(x, Q^2)$ is the gluon distribution function, and
\be
g_{L,R}^2 \equiv \left(g^u_{L,R}\right)^2 
+ \left(g_{L,R}^d\right)^2 ~.
\ee
The $\CF_3$ structure functions at NLO  
does not get a contribution from 
scattering off the gluon sea, and has a 
similar form for the charged- and
neutral-currents,
\be
\delta \CF_3^{\rm C,N} = 
-\frac{2\alpha_s}{3\pi}  \int_x^1 dz
\left(1+\frac{1}{z}\right)
\CF_{3\, {\rm LO}}^{\rm C,N}\left(\frac{x}{z}, Q^2\right) ~.
\label{f3-nlo}
\ee
These expressions apply to the $\overline{\nu}_\mu$-nucleus 
processes as well, with the only difference that 
the  $q$ and $\overline{q}$ distributions have to be interchanged in the 
expressions for the leading-order structure functions given in
Eqs.~(\ref{CC-structure-functions}) and (\ref{NC-structure-functions}).

Although corrections due to electromagnetic radiation, 
electroweak loops, target mass, and fermion masses, 
are important for the lowest-order cross sections, as discussed in Section 2,
they can be neglected in the computation of the order-$\alpha_s$ corrections.
Formally, they represent higher-order terms in the expansion (\ref{corr}).
For example, the parton level processes $\nu g \rightarrow \nu c\overline{c}$
and $\nu_\mu g \rightarrow \mu^- c\overline{s}$ are suppressed at
small $Q^2$, which is an order $(\alpha_s/\pi)(m_c^2/Q^2)$
effect.

\section{Total Cross Sections for $\nu$-Fe Scattering}
\setcounter{equation}{0}

In this Section we derive some analytical, approximate expressions
for the total cross sections in neutrino deep-inelastic scattering.
We begin by expanding the gauge boson propagator 
in powers of $Q^2/M_{W,Z}^2$, and use Eq.~(\ref{qsquare}): 
\be
\frac{1}{ \left(1 + Q^2/M_{W,Z}^2\right)^2 } 
\approx 1 - \frac{4xyM_N E_\nu}{M_{W,Z}^2} 
+ O\left(\frac{(xyM_N E_\nu)^2}{M_{W,Z}^4} \right)~.
\label{expand}
\ee
This enables us to take advantage of the following identity
\be
\int_0^1 dx \; x^{n-1} \int_x^1 \frac{dz}{z} \; 
f(z) q\left(\frac{x}{z}\right) = 
q^{(n)}  \int_0^1 dz \; z^{(n-1)} f(z) ~,
\ee
where $f(z)$ is any non-singular function, and 
\be
q^{(n)} \equiv \int_0^1 dx \; x^{n-1} q(x) 
\ee
is the $n$th moment of the $q(x)$ parton distribution.

In what follows we will keep only the 
leading term of the expansion shown in Eq.~(\ref{expand}).
Furthermore, when computing 
the $\delta \CF_i$  corrections to the structure functions, 
given in Eqs.~(\ref{f1-nlo}) and (\ref{f3-nlo}),
the evolution of the quark and gluon PDF's, 
$q_j(x,Q^2)$ and $g(x,Q^2)$, may be approximated by 
taking the PDF's at the average $Q^2$, labeled $\overline{Q^2}$, as long 
as the range of $Q^2$ is not too large.
The error on the cross section, due to this approximation of 
the NLO QCD corrections,
is of the order of $\alpha_s^2(\overline{Q^2}) \ln (Q^2/\overline{Q^2})$.

As a result, the integration over $x$ and $y$ of the differential
cross-sections given in  Eq.~(\ref{differential}) yields
\be
\sigma^{\rm C,N}\left( \nu_\mu {\rm Fe} \right)
= \frac{M_N E_\nu G_{\rm F}^2}{6\pi}
\left(\CF_1^{{\rm C,N}(2)}
+ 3 \CF_2^{{\rm C,N}(2)}
+ 2\CF_3^{{\rm C,N}(2)}\right) ~.
\label{total}
\ee
The second moments of the structure functions are given by 
\bear
\CF_i^{{\rm C,N}(2)} & = & \int_0^1 dx \, x \; \CF_i^{\rm C,N} (x) 
\nonumber \\ [1em] 
 & = & \CF_{i \, {\rm LO}}^{{\rm C,N}(2)} + \delta \CF_{i}^{{\rm C,N}(2)}
+ O\left(\frac{Q^2}{M_{W,Z}^2} \, , \,
\frac{\alpha}{\pi\sin^2\theta_W} \, , \, 
\frac{M_N^2}{Q^2} \, , \, 
\frac{m_c^2}{Q^2}\right)
~,
\eear
for $i=1,2,3$. The second moments of the lowest-order structure
functions, $\CF_{i \, {\rm LO}}^{{\rm C,N}(2)}$, are obtained simply by 
taking the second moments of the PDF's in 
Eqs.~(\ref{CC-structure-functions}) and (\ref{NC-structure-functions}),

The second moments of the $\delta \CF_{i}^{{\rm C,N}}$ corrections 
to the structure functions are given by 
\bear 
&&
\delta \CF_1^{{\rm C}(2)} =
- \frac{4\alpha_s}{9\pi} \left[ \CF_{1\, {\rm LO}}^{{\rm C}(2)}
+ \frac{3}{2} g^{(2)} \right] ~,
\nonumber \\ [1em] &&
\delta \CF_1^{{\rm N}(2)} = 
- \frac{4\alpha_s}{9\pi} \left[ \CF_{1\, {\rm LO}}^{{\rm N}(2)}
+ \frac{3}{2}\left(g_L^2 + g_R^2\right) g^{(2)} \right] ~,
\nonumber \\ [1em] &&
\delta \CF_3^{{\rm C,N}(2)} = -\frac{5\alpha_s}{9\pi} 
\CF_{3\, {\rm LO}}^{{\rm C,N}(2)} ~,
\label{moments1-nlo}
\eear
where $g^{(2)}$ is the second moment of the gluon distribution function.
Recall that these results are obtained in the DIS scheme,
where $\delta \CF_2^{{\rm C,N}(2)} = 0$.

\section{Estimate of the neutral-current to charged-current event ratio}
\setcounter{equation}{0}

Although an analysis of the data involving the NLO QCD corrections
to the differential cross sections [Eqs.~(\ref{f1-nlo}) and (\ref{f3-nlo})]
is required for a precise determination of the shift in $\sin^2\!\theta_W$, 
we now show that it is also possible to estimate theoretically this shift. 

\subsection{General results}

The approximate expressions that we obtained 
for the total cross sections, Eq.~(\ref{total}), have the same $E_\nu$-dependence for
both the neutral-current and charged-current events. Therefore,
the ratio of neutral- to charged-current events
is independent of the neutrino flux, and is given by 
the ratio of total cross sections.
At leading order in $\alpha_s$, $\alpha$, and the various mass ratios, 
this is
\bear
R_0^\nu & = & \frac{ 2\CF_{1\, {\rm LO}}^{{\rm N}(2)}
+ \CF_{3\, {\rm LO}}^{{\rm  N}(2)} }
{2\CF_{1\, {\rm LO}}^{{\rm C}(2)}
+ \CF_{3\, {\rm LO}}^{{\rm C}(2)} }
\nonumber \\ [1em] 
 & = & g_L^2 + r g_R^2 - 
\left( {g_L^u}^2 - \frac{{g_R^d}^2}{3} \right) \frac{q_- }{q_0 } 
+ \left( \frac{{g_L^d}^2}{3} - {g_R^u}^2 \right) \frac{ \overline{q}_- }{ q_0 } 
~,
\label{zero-order}
\eear
where we have introduced two linear combinations of second moments,
\bear &&
q_0 \equiv  d^{(2)} + s^{(2)} + \frac{1}{3}\left(\overline{u}^{(2)} 
+ \overline{c}^{(2)}\right) ~,
\nonumber \\ [1em] &&
q_- \equiv d^{(2)} - u^{(2)} + s^{(2)} -  c^{(2)} ~.
\eear
The ratio $r$ of the total cross sections 
for the $\bar{\nu}$Fe and $\nu$Fe charged-current processes at leading order,
is simply
\be
r = \frac{ \overline{q}_0 }{ q_0 } ~.
\ee

The  ratio of neutral- to charged-current events 
is changed by the NLO QCD effects to
\be
R^\nu = \frac{ \sigma^{\rm N}\left( \nu_\mu {\rm Fe} \right) }
{ \sigma^{\rm C}\left( \nu_\mu {\rm Fe} \right) }
= R_0^\nu + \delta R^\nu_1 + \delta R^\nu_3 + O\left(\frac{Q^2}{M_{W,Z}^2} \, , \,
\frac{\alpha}{\pi\sin^2\theta_W} \, , \, 
\frac{M_N^2}{Q^2} \, , \, 
\frac{m_c^2}{Q^2}\right) ~.
\ee
The shift in $R^\nu$ from order $\alpha_s$ corrections to $\CF_i$, $i = 1,3$, 
follows from Eq.~(\ref{total}):
\be
\delta R^\nu_i = c_i
\frac{ \delta \CF_i^{{\rm N}(2)} - R_0^\nu \delta \CF_i^{{\rm C}(2)} }
{2 \CF_{1\, {\rm LO}}^{{\rm C}(2)} + \CF_{3\, {\rm LO}}^{{\rm C}(2)} }  ~,
\ee
where $c_1 = 1/2$ and $c_3 = 1$.

The above equation, along with the expressions for the second moments of the 
leading-order PDF's 
[see Eqs.~(\ref{CC-structure-functions}) and (\ref{NC-structure-functions})] 
and their NLO corrections given in 
Eq.~(\ref{moments1-nlo}) lead to an analytic formula for the shift in $R^\nu$ 
in terms of measured quantities. 
This involves only two more linear combinations of second moments:
\bear && 
q_1 \equiv  d^{(2)} + s^{(2)} + \overline{u}^{(2)} + \overline{c}^{(2)} +
\frac{3}{4} g^{(2)} ~,
\nonumber \\ [1em] &&
q_3 \equiv d^{(2)} - \overline{u}^{(2)} + s^{(2)} - \overline{c}^{(2)} ~.
\eear
The final result is 
\bear \hspace*{-.3em}
\delta R^\nu_1 & = & - \frac{2 \alpha_s}{27 \pi} \left\{
g_R^2 (1-r) \frac{ q_1 }{ q_0 } - \frac{ q_- }{ q_0 } \left[ {g_L^u}^2
+ {g_R^u}^2 - \left( {g_L^u}^2 -  \frac{{g_R^d}^2}{3} \right)
\frac{ q_1 }{ q_0 } \right] \right.
\nonumber \\ [.7em] && \hspace*{.3em} \left.
+ \frac{ \overline{q}_- }{ q_0 } \left[ {g_L^d}^2
+ {g_R^d}^2 - \left( \frac{{g_L^d}^2}{3} - {g_R^u}^2 \right)
\frac{ q_1 }{ q_0 } \right] \right\} ~,
\nonumber \\ [1.5em] \hspace*{-.3em}
\delta R^\nu_3 & = & \frac{5 \alpha_s}{27 \pi} \left\{
g_R^2 (1+r) \frac{ q_3 }{ q_0 } 
+ \frac{ q_- }{ q_0 } \left[ {g_L^u}^2
- {g_R^u}^2 - \left( {g_L^u}^2 - \frac{ {g_R^d}^2}{3} \right)
\frac{ q_3 }{ q_0 } \right] \right.
\nonumber \\ [.7em] && \hspace*{.3em} \left.
+ \frac{ \overline{q}_- }{ q_0 } \left[ {g_L^d}^2
- {g_R^d}^2 + \left( \frac{{g_L^d}^2}{3} - {g_R^u}^2 \right)
\frac{ q_3 }{ q_0 } \right] \right\}  ~.
\label{final}
\eear

For $\overline{\nu}$Fe scattering, the ratio of neutral- 
to charged-current events at leading order, $R_0^{\overline{\nu}}$, 
is given by the right-hand side of Eq.~(\ref{zero-order}) with the following
substitutions: $r \rightarrow 1/r$, $q_0 \rightarrow \overline{q}_0$,
$q_- \leftrightarrow \overline{q}_-$. This is shifted 
at NLO by QCD effects by 
$\delta R^{\overline{\nu}}_1 + \delta R^{\overline{\nu}}_3$, where 
$\delta R^{\overline{\nu}}_{1,3}$ are obtained from 
Eq.~(\ref{final}) by performing the same 
substitutions as above, and in addition $q_1 \rightarrow \overline{q}_1$,
$q_3 \rightarrow \overline{q}_3$.

\subsection{Origin of the $\sin^4\theta_W$ suppression}

Before evaluating the size of the NLO corrections given 
in Eq.~(\ref{final}), there is an important observation to be made.
In the ``enhanced isospin symmetry'' limit, where 
\be
d^{(2)} = u^{(2)} \, , \;\; 
\overline{d}^{(2)} = \overline{u}^{(2)} \, , \;\; 
s^{(2)} = c^{(2)} \, , \;\; 
\overline{s}^{(2)} = \overline{c}^{(2)} \, , 
\label{enhanced}
\ee
so that $q_- = \overline{q}_- = 0$, Eq.~(\ref{final}) implies 
that $\delta R^\nu_{1,3}$ are parametrically of the order of
$g_R^2\alpha_s/\pi$. Given that 
\be
g_R^2 = (5/9) \sin^4\theta_W \approx 2.76 \times 10^{-2} ~,
\ee
the NLO QCD corrections to $R^\nu$ are suppressed by a factor 
of approximately 30 compared to the naive expectation 
of $\alpha_s/\pi$. It is therefore interesting to understand 
the origin of this suppression. 

To this end, notice that in the limit where the quark masses
are ignored, the cross section for the neutral-current process 
can be written as a sum 
of cross sections for neutrino scattering off left- and right-handed
quarks:
\be
\sigma^{\rm N}\left( \nu_\mu {\rm Fe} \right)
= \sigma^{\rm N}_{0 L} + \sigma^{\rm N}_{0 R} +
\delta \sigma^{\rm N}_{L} 
+ \delta \sigma^{\rm N}_{R} ~,
\ee 
where the subscript 0 refers to the leading order terms, and
$\delta \sigma$ are the QCD corrections.
If the enhanced isospin symmetry were exact, then
\be
\frac{\sigma^{\rm N}_{0 L}}{\sigma^{\rm C}_{0}} = 
\frac{\delta \sigma^{\rm N}_{L}}{\delta \sigma^{\rm C}}  = g_L^2 ~,
\ee
so that 
\be
R^\nu = \frac{\sigma^{\rm N}_{0 L}}{\sigma^{\rm C}_{0}}
+ \frac{\sigma^{\rm N}_{0 R} + \delta \sigma^{\rm N}_{R} }
{\sigma^{\rm C}_{0} + \delta \sigma^{\rm C}} ~.
\ee
This equation shows that the QCD corrections
to $R^\nu$ would vanish {\it to all orders} 
if the neutral-current
involving the right-handed quarks were not present
[in the limit where the quarks are massless and the 
enhanced isospin symmetry, Eq.~(\ref{enhanced}) is exact].
The factor of $g_R^2$ is a consequence of this fact.

In reality isospin symmetry is broken due to the different
number of neutrons and protons in the target, as well as by the 
quark mass differences and electromagnetic interactions. 
Therefore, in addition to the terms of order $g_R^2\alpha_s/\pi$,
$R^\nu$ gets corrections of order $(q_-/q_0) \alpha_s/\pi$,
as can be seen in Eq.~(\ref{final}).
For an approximately isoscalar target such as
iron, the terms of order $g_R^2\alpha_s/\pi$ dominate, albeit 
by a small margin, as discussed in the next subsection.

\subsection{Size of the corrections to $R^\nu$ and $R^{\overline{\nu}}$}

The nine second moments, $u^{(2)},d^{(2)},s^{(2)},
c^{(2)}$,$\overline{u}^{(2)}, \overline{d}^{(2)}, \overline{s}^{(2)}, 
\overline{c}^{(2)}$ and $g^{(2)}$, are given by an average over the 
second moments of the nucleon PDF's inside the iron nucleus, 
with corrections due to nuclear interactions. 
They are evaluated at an average $Q^2$. For NuTeV,
the average value for $Q^2$ is 25.6 GeV$^2$ for the $\nu_\mu$ beam and 
15.4 GeV$^2$ for the $\overline{\nu}_\mu$ beam. 
We choose $\overline{Q^2}$ to be around 20 GeV$^2$.

The PDF's used in the NuTeV analysis come from a fit 
to the charged-current differential
cross sections measured by the CCFR experiment \cite{Yang:2000ju}
with the same iron target. The fit and the 
Monte Carlo simulation used for extracting $\sin^2\theta_W$ 
employ the same cross section model, which is described in 
Ref.~\cite{Zeller:2002he}.
At $\overline{Q^2} = 20$ GeV$^2$,
the fit gives the following values for the second 
moments  \cite{sam}:
$u^{(2)} \approx 0.196$,
$d^{(2)}\approx 0.204$,
$\overline{u}^{(2)}\approx \overline{d}^{(2)}\approx 0.032$,
$s^{(2)}\approx \overline{s}^{(2)}\approx 0.013$,
$c^{(2)}\approx \overline{c}^{(2)}\approx 0.006$,  
and $g^{(2)}\approx 0.498$.
This fit assumed $s(x, Q^2) = \overline{s}(x, Q^2)$, 
 $c(x, Q^2) = \overline{c}(x, Q^2)$, and isospin symmetry
in the sense that the only difference between the 
$u$ and $d$ distributions is due to the different number 
of protons and neutrons in the iron nucleus. 

An asymmetry of order a few percent 
between the $s$ and $\overline{s}$ distributions, 
and isospin-breaking effects, 
due to the up-down quark mass splitting and electroweak interactions, 
expected to be of order $(m_d - m_u)/\Lambda_{\rm QCD}$, 
{\it i.e.} also a few percent, would be important for the leading 
order $R_0^{\nu,\overline{\nu}}$ ratios \cite{Davidson:2001ji}, 
but can be neglected in the estimate 
of the NLO corrections.
Also, the shifts $\delta R_{1,3}^{\nu,\overline{\nu}}$ are only mildly 
sensitive to the choice of a different set of PDF's. The main reason 
is that only five independent combinations of second moments appear in 
Eqs.~(\ref{final}):
$r\approx 0.49$, $q_1/q_0 \approx 2.74$, 
$q_3/q_0\approx 0.78$, $q_-/q_0 \approx 0.07 $, 
$\overline{q}_-/q_0\approx 0.03$.
The other relevant combinations of second moments can be expressed in terms of these.
For example,
\bear &&
\frac{\overline{q}_1}{\overline{q}_0} = \frac{q_1 - q_- + \overline{q}_-}{r q_0} ~,
\nonumber \\ [.7em] &&
\frac{\overline{q}_3}{\overline{q}_0} = \frac{-q_3 + q_- + \overline{q}_-}{r q_0} ~.
\eear
Note that $q_-/q_0$ and $\overline{q}_-/q_0$ have values comparable with 
$g_R$, and therefore the terms proportional with 
$(q_-/q_0) \alpha_s/\pi$ and $(\overline{q}_-/q_0) \alpha_s/\pi$
cannot be neglected in Eq.~(\ref{final}). Nonetheless, these terms
are also multiplied by factors of the order of ${g_L^u}^2\approx 0.12$ 
and ${g_L^d}^2\approx 0.18$, so that the isospin symmetric 
corrections of order $g_R^2\alpha_s/\pi$ dominate.


For $\alpha_s = 0.2$ and $\sin^2\!\theta_W = 0.2227$
we obtain the following values for the shifts 
in $R^\nu$ due to NLO QCD corrections to 
the $\CF_1$ and $\CF_3$ structure functions:
\bear 
&&
\delta R^\nu_1 \approx -2.5 \times 10^{-4} ~,
\nonumber \\ [.7em] &&
\delta R^\nu_3 \approx 4.6 \times 10^{-4} ~.
\label{shift-nu}
\eear
In the case of the $\overline{\nu}$ beam, the results are
\bear
&&
\delta R^{\overline{\nu}}_1 \approx 6.1 \times 10^{-4} ~,
\nonumber \\ [.7em] &&
\delta R^{\overline{\nu}}_3 \approx -9.9 \times 10^{-4} ~.
\label{shift-nubar}
\eear
These corrections are of the order of the standard deviations 
quoted by the NuTeV collaboration \cite{Zeller:2001hh} 
for the measured $R^\nu_{\rm exp}$ and $R^{\overline{\nu}}_{\rm exp}$: 
$7 \times 10^{-4}$ and $16 \times 10^{-4}$, respectively.
Note though that the measured quantities ($R^\nu_{\rm exp}$
and $R^{\overline{\nu}}_{\rm exp}$) are ratios of the numbers 
of short and long events observed in the NuTeV detector, and therefore 
differ from the ratios of neutral- and charged-current events
($R^\nu$ and $R^{\overline{\nu}}$) due to the 
experimental cuts, backgrounds and detector acceptance.
A discussion of  these effects, 
albeit primarily in the context of QCD corrections to 
$R_{\rm PW}$, is given in Ref.~\cite{McFarland:2003jw}.

Comparing our results given in 
Eq.~(\ref{shift-nu}) and (\ref{shift-nubar}) with the numerical 
results given in Ref.~\cite{Kretzer:2003iu} we observe that 
the size of the effect is of the same order of magnitude, but 
the sign of $\delta R^\nu = \delta R^\nu_1 + \delta R^\nu_3$ 
is opposite. The various approximations that we have employed 
in obtaining the analytical expression for $\delta R^\nu$,
such as ignoring the charm mass and the evolution of the PDF's,
which introduce errors of the order of
$(\alpha_s/\pi)(m_c^2/\overline{Q^2})$ and  
$\alpha_s^2(\overline{Q^2}) \ln (Q^2/\overline{Q^2})$, respectively,
do not seem to be sufficient to account for this difference.
It remains to be seen whether the effect of
the hadronic energy cut used in Ref.~\cite{Kretzer:2003iu}
is large enough to explain the difference \cite{sam}.

\section{Theoretical estimate of the shift in $\sin^2\!\theta_W$}
\setcounter{equation}{0}

The approximate results for the shifts in $R^\nu$
and $R^{\overline{\nu}}$ obtained in the previous section 
should in principle allow an estimate of the 
corrections to the value of $\sin^2\!\theta_W$ determined by the 
NuTeV collaboration. In practice, however, there are several elements 
in the NuTeV analysis that make a theoretical estimate somewhat 
problematic. Here we point out a few complications.

\subsection{Relation between $\sin^2\!\theta_W$ and $R^\nu$,
$R^{\overline{\nu}}$}

The NuTeV analysis includes a phenomenological description of the 
so-called longitudinal structure function, which changes the relation between
the $\CF_1$ and $\CF_2$ structure functions. Effectively, this procedure 
approximately accounts for the QCD corrections to $\CF_1$. 
We will therefore consider only the impact of the QCD corrections to $\CF_3$, 
which lead to the values for $\delta R^\nu_3$ and $\delta R^{\overline{\nu}}_3$
given in Eqs.~(\ref{shift-nu}) and (\ref{shift-nubar}). 

Naively, the shift in $\sin^2\!\theta_W$ due to a shift in 
the predicted value of $R^\nu$
can be derived immediately from the expression of $R_0^\nu$
given in Eq.~(\ref{zero-order}):
\be
\delta \sin^2\!\theta_W \approx \frac{\delta R^\nu_3}
{ 1 - (10/9)(1+r)\sin^2\!\theta_W } \approx 0.7 \times 10^{-3}~.
\ee
However, various effects change this relation.
These include a ``cross-talk'' between the 
charged- and neutral-current events, experimental cuts, 
and the corrections to the structure functions 
listed in Eq.~(\ref{corr}).
The NuTeV analysis has computed these effects using a 
Monte Carlo simulation.
Note that  $\delta R^\nu_3$
and $\delta R^{\overline{\nu}}_3$ can be viewed as approximate 
shifts in the results for $R^\nu$ and $R^{\overline{\nu}}$
given by the Monte Carlo simulation used by NuTeV. 
The relation between these shifts and the shift
in $\sin^2\!\theta_W$ is given in section 8
of Ref.~\cite{Zeller:2002he}:
\be
\delta \sin^2\!\theta_W = \frac{1}{b}\left( \delta R^\nu_3
- a \, \delta R^{\overline{\nu}}_3 \right)~.
\ee
For the fit reported in the NuTeV result \cite{Zeller:2001hh},
where the charm mass is constrained, $a = 0.249$ and $b = 0.617$,
giving $\delta \sin^2\!\theta_W \approx  1.1 \times 10^{-3}$,
which is an increase of about $0.7\sigma $.
For the fit without constraints, $a = 0.453$ and $b = 0.612$,
and the increase in $\sin^2\!\theta_W$ is close to $1\sigma$.
Thus, the inclusion of the corrections to $\CF_3$ alone 
tend to increase the deviation from the Standard Model. 

\subsection{QCD corrections to the parton distributions} 

The $Q^2$ dependence of the PDF's is an effect of order 
$\alpha_s(\overline{Q^2})/\pi \ln (Q^2/\overline{Q^2})$, where 
$\overline{Q^2}$ is an average value for $Q^2$.
The NuTeV collaboration has approximated the $Q^2$ dependence 
by the Buras-Gaemers evolution\cite{Buras:1977yj}. 
Using the exact QCD evolution could modify the values 
derived from the CCFR data 
of the PDF's at our reference point of $Q^2 = 20$ GeV$^2$.
We will not attempt here to estimate this effect.
We only mention that this leads to a correction 
to $\sin^2\!\theta_W$ that is independent of 
the one given in Eq.~(\ref{final-sw}). Only at order 
$g_R^2 \alpha_s^2(\overline{Q^2})/\pi^2 \ln (Q^2/\overline{Q^2})$
the two corrections become correlated.

The PDF's used by NuTeV collaboration
are extracted from a fit of the differential cross sections 
to the $\nu$ and $\overline{\nu}$ charged-current CCFR data.
The inclusion of order $\alpha_s$ terms in the cross sections changes 
the fit. The shifts in the quark PDF's lead to 
corrections of order $\alpha_s$ to the $r$ ratio that enters in the 
expression for $R_0^\nu$ given in Eq.~(\ref{zero-order}).
Therefore, we expect additional corrections to $\sin^2\!\theta_W$
of order $g_R^2 \alpha_s/\pi$, that may change the result by a factor 
of order unity and unknown sign.

\section{Conclusions}

We have presented an analysis of the $O(\alpha_s)$ radiative corrections
to the ratios of neutral- and charged-current cross sections, 
$R^\nu$ and  $R^{\bar{\nu}}$. We have shown that these effects are smaller
than the $O(\alpha_s/\pi)$ one might expect a priori, because of a 
suppression factor of  $\sin^4 \theta_W$ in the dominant contribution. 
On the other hand, the effects 
turn out to be of the same order as the 1-$\sigma$ error in the experimental
results of NuTeV. 

Our results indicate the importance of a full NLO analysis
of the NuTeV data, which would include the NLO QCD corrections 
to the cross sections [see Eqs.~(\ref{f1-nlo}) and (\ref{f3-nlo})] as well 
as the QCD evolution of the PDF's, in both 
the Monte Carlo simulation used for determining $\sin^2 \theta_W$ and 
the fit to the charged-current data used for extracting the PDF's. 
In addition, our results will provide a simple check when such an
analysis is performed. 

It is important to keep in mind that the NLO QCD corrections
discussed here are independent 
at this order of the corrections 
discussed in Refs.~\cite{Zeller:2002du, Martin:2003sk, cteq}, which 
require a refit of the data that  allows
{\it both} a strange asymmetry and a violation of isospin symmetry.

\bigskip\medskip

{\bf Acknowledgments}: We are grateful to K.~McFarland and G.~P.~Zeller
for many useful discussions on the NuTeV data analysis.
We  would like to thank U.~Baur, P.~Gambino, 
S.~Kretzer, F.~Olness, M.~H.~Reno
and W.-K.~Tung for helpful conversations.
Fermilab is operated by University Research Association, Inc., under
contract DE-AC02-76CH03000.

 
\vfil \end{document}